\input{style/arxiv-general.cfg}
\documentclass[aoas,MSNbibl,nameyear,seceqn,dvips]{arximspdf}
\makeatletter
   \@ifpackageloaded{graphicx}{}{\usepackage{graphicx}}
\makeatother
\usepackage{textcomp}

\doi{10.1214/14-AOAS797}
\volume{9}
\issue{1}
\pubyear{2015}
\firstpage{383}
\lastpage{401}
\docsubty{FLA}

\makeatletter
\renewcommand{\mid}{|}
\newcommand{\cT}{\mathcal{T} }
\newcommand{\var}{\operatorname{Var}}
\makeatother

\begin{document}
\begin{frontmatter}

\title{A Bayesian regression tree approach to identify the~effect of
nanoparticles' properties on toxicity~profiles\thanksref{T1}}
\runtitle{Bayesian regression trees for nanotoxicity profiles}

\begin{aug}
\author[A]{\fnms{Cecile}~\snm{Low-Kam}\corref{}\ead[label=e1]{clowkam@gmail.com}},
\author[A]{\fnms{Donatello}~\snm{Telesca}\ead[label=e2]{dtelesca@ucla.edu}},
\author[B]{\fnms{Zhaoxia}~\snm{Ji}\ead[label=e3]{zji@cnsi.ucla.edu}},
\author[B]{\fnms{Haiyuan}~\snm{Zhang}\ead[label=e4]{zhangh@ucla.edu}},
\author[C]{\fnms{Tian}~\snm{Xia}\ead[label=e5]{txia@ucla.edu}},
\author[D]{\fnms{Jeffrey~I.}~\snm{Zink}\ead[label=e6]{zink@chem.ucla.edu}}
\and
\author[E]{\fnms{Andre~E.}~\snm{Nel}\ead[label=e7]{ANel@mednet.ucla.edu}}
\runauthor{C. Low-Kam et al.}
\affiliation{University of California, Los Angeles}
\address[A]{C. Low-Kam\\
D. Telesca\\
Department of Biostatistics\\
University of California\\
Los Angeles, California 90095\\
USA\\
\printead{e1}\\
\phantom{E-mail: }\printead*{e2}}
\address[B]{Z. Ji\\
H. Zhang\\
California NanoSystems Institute\hspace*{27pt}\\
University of California\\
Los Angeles, California 90095\\
USA\\
\printead{e3}\\
\phantom{E-mail: }\printead*{e4}}
\address[C]{T. Xia\\
Division of NanoMedicine\\
Department of Medicine\\
University of California\\
Los Angeles, California 90095\\
USA\\
\printead{e5}}
\address[D]{J. I. Zink\\
Department of Chemistry \& Biochemistry\\
University of California\\
Los Angeles, California 90095\\
USA\\
\printead{e6}}
\address[E]{A.~E. Nel\\
California NanoSystems Institute\\
University of California\\
Los Angeles, California 90095\\
USA\\
and\\
Division of NanoMedicine\\
Department of Medicine\\
University of California\\
Los Angeles, California 90095\\
USA\\
\printead{e7}}
\end{aug}
\thankstext{T1}{Supported by U.S. Public Health
Service Grant U19 ES019528
(UCLA Center for Nanobiology and Predictive Toxicology) and 
supported
by the National Science Foundation and the Environmental Protection
Agency under Cooperative
Agreement Number DBI-0830117.}

%
\received{\smonth{2} \syear{2014}}
%
\revised{\smonth{11} \syear{2014}}

%
\begin{abstract}
We introduce a Bayesian multiple regression tree model to characterize
relationships between
physico-chemical properties of nanoparticles and their in-vitro
toxicity over multiple doses and times of exposure.
Unlike conventional models that rely on data summaries, our model
solves the low sample size
issue and avoids arbitrary loss of information by combining all
measurements from a general exposure experiment across doses, times of
exposure, and replicates.
The proposed technique integrates Bayesian trees for modeling threshold
effects and interactions,
and penalized B-splines for dose- and time-response surface smoothing.
The resulting posterior
distribution is sampled by Markov Chain Monte Carlo. This method
allows for inference on a number of quantities of potential interest to
substantive nanotoxicology,
such as the importance of physico-chemical properties and their
marginal effect on toxicity.
We illustrate the application of our method to the analysis of a
library of 24 nano metal oxides.
\end{abstract}

%
\end{frontmatter}

\section{Introduction}

The increasing use of engineered nanomaterials (ENM) in hundreds of
consumer products has recently raised concern about their potential
effect on the environment and human health in particular. In
nanotoxicology, in vitro dose-escalation assays describe how cell lines
or simple organisms are affected by increased exposure to
nanoparticles. These assays help determine \mbox{hazardous} materials and
exposure levels.
Standard dose-escalation studies are sometimes completed by more
general exposure escalation protocols, where a biological outcome is
measured against both increasing concentrations and durations of exposure.
Cost and timing issues usually only allow for a small number of
nanoparticles to be comprehensively screened in any
study. Therefore, both one- and two-dimensional escalation experiments
are often characterized by small sample sizes.
Furthermore, data exhibits natural clusters related to varying levels
of nanoparticles bio-activity.
The two case studies presented in Section~\ref{sectionappli} provide an
overview of the structure of typical data sets obtained with both
experimental protocols.

Beyond dose-response analysis, nanomaterial libraries are also designed
to investigate how
a range of physical and chemical properties (size, shape, composition,
surface characteristics) may
influence ENM's interactions with biological systems.
The nano-informatics literature reports several Quantitative
Structure--Activity Relationship (QSAR) models.
This exercise is conceived as a framework for predictive toxicology,
under the assumption that nanoparticles with similar properties are
likely to have similar effects.
Most of existing QSAR models summarize or integrate experimental data
across times, doses and replicates as a preprocessing step, before
applying traditional data mining or statistical algorithms for regression.
For example, \citet{rong} use a modified Student's $t$-statistic to
discretize outputs in two classes (toxic or nontoxic) and a logistic
regression model to relate toxicity to physico-chemical variables.
\citet{zhang} use the area under the dose-response curve as a
global summary of toxicity and they model dependence on predictors via
a regression tree.
Both approaches, while reasonably sensible, ignore the uncertainty
associated with data summaries and can lead to unwarranted conclusions
as well as unnecessary loss of information.
\citet{patel13} summarize toxicity profiles using a new definition
of toxicity, called \emph{the probability of toxicity}, which is
defined as a linear function of nanoparticle physical and chemical properties.
While this last approach solves the issue of uncertainty propagation,
it still makes it impossible to predict full dose-response curves from
nanoparticle characteristics.
Moreover, the use of regression trees is inherently appealing, as they
are able to model nonlinear effects and interactions
without compromising interpretation. We aim to extend regression tree
models to account for structured multivariate outcomes, defined as
toxicity profiles of
nanoparticles, measured over a general exposure escalation domain.

Multivariate extensions of the regression tree methodology have been
proposed by \citet{segal92}. In this paper, the original
tree-building algorithm of \citet{cart84} is modified to handle
multivariate responses for commonly used covariance matrices, such as
independence or autoregressive structures.
\citet{death02} proposes a similar method for an independent
covariance structure. \citet{yu99} develop regression tree models
for functional data, by representing each individual response as a
linear combination of spline basis functions and using the estimated
splines coefficients in multivariate regression trees.
An alternative for longitudinal responses consists of combining a tree
model and a linear model: \citet{sela10} replace the fixed effects
of the traditional linear mixed effects model by a regression tree. The
linear random effects are unchanged. \citet{yu10} fit a
semi-parametric model, containing a linear part and a tree part, for
multivariate outcomes in genetics. The linear part is used to model
main effects of some genetic or environmental exposures. The
nonparametric tree part approximates the joint effect of these
exposures. Finally, \citet{galimberti02} develop regression tree
models for longitudinal data with time-dependent covariates. In this
setting, measures for the same individual can belong to different
terminal nodes.

Other extensions of standard regression trees include Bayesian
approaches, where tree parameters become random variables. \citet
{chipman97} introduce a Bayesian regression tree model for univariate
responses. The method is based on a prior distribution and a
Metropolis--Hastings algorithm which generates candidate trees and
identifies the most promising ones. This methodology has since been
extended to so-called \emph{treed} models, where a parametric model is
fitted in each terminal node [\citet{chipman03}],
to a sum-of-trees model [\citet{bart}], and to incorporate spatial
random effects for merging data sets [\citet{muller}].
\citet{TGP} model nonstationary spatial data by combining Bayesian
regression trees and Gaussian processes in the leaves. This approach is
extended to the multivariate Gaussian process with separable covariance
structure in \citet{MTGP}.

Building on previous contributions, we propose a new method to analyze
the relationship between nanoparticles physico-chemical properties and
their toxicity in exposure escalation experiments. We extend the
Bayesian methodology of \citet{chipman97} to allow for dose- and
time-response kinetics in terminal nodes.
Our work is closely related to the methodology introduced in \citet
{MTGP}. However, our model is specifically adapted to exposure
escalation experiments, as observations for the same nanoparticle at
different doses and times cannot fall in separate leaves of the tree.
Therefore, the binary splits of the tree only capture structure
activity relationships instead of the general increase of toxicity with
exposure.

A global covariance structure accounts for correlation between
measurements at different doses and times for the same nanoparticle.
Our approach is able to model nonlinear effects and potential
interactions of physico-chemical properties without making parametric
assumptions about toxicity profiles. It also addresses the issues
associated with conventional QSAR models by combining evidence across
measurements for all doses and times in a general experimental design.
The proposed model is particularly versatile, as it provides scores of
importance for physico-chemical properties and visual assessment of the
marginal effect of these properties on toxicity.

The rest of this paper is organized as follows: Section~\ref
{sectionmodel} describes the regression model for dose-response data
and Section~\ref{sectionprior} describes the corresponding prior model.
The resulting posterior distribution and the associated MCMC algorithm
are presented in Section~\ref{sectionalgo}.
The model is extended to the case of dose- and time-response surfaces
in Section~\ref{sectiongeneralcase}. The method is applied to a library
of 24 metal oxides in Section~\ref{sectionappli} and Section~\ref
{sectiondiscu} concludes this paper with a discussion.

\section{Regression tree formulation}\label{sectionmodel}

\subsection{Sampling model}
We first consider the case of a typical dose escalation experiment, where
a biological outcome is measured over a protocol of increased
nanoparticle concentration. This case will be expanded in Section~\ref
{sectiongeneralcase} to include more general exposure escalation designs.

Let $y_{ik}(d)$ denote a real-valued response associated with exposure
to nanoparticle $i$ and replicate $k$ at dose $d$, for $i \in\{1,
\ldots, I\}$, $k \in\{1, \ldots, K\}$ and $d\in[0, D]$. We assume that
$y$ has been appropriately normalized and purified of experimental
artifacts. For the two case studies of Section~\ref{sectionappli},
normalization was performed for each tray by subtracting a baseline
mean response, measured in control wells where cells were not exposed
to any nanoparticle. After normalization, we indeed assume 
independence between wells exposed to different nanomaterials on the
same tray.
Current experimental protocols only allow for the observation of the
outcome $y$ as it varies in association with a discrete prescription of
dose-escalation. However, for notational convenience and without loss
of generality, we maintain that $y$ shall be observed for any dose
level $d$ ranging from no exposure $(d=0)$ to a maximal nanoparticle
concentration level $(d=D)$.
Let also $\mathbf{x}_i'=(x_{i1},\ldots,x_{ip})$ be a $p$-dimensional
vector of continuous physico-chemical characteristics or predictors
associated to nanoparticle $i$. We assume
%
\begin{eqnarray}
y_{ik}(d) & = & f(\mathbf{x}_i,d) + \varepsilon_{ik}(d),
\label{eqmodel1}
\end{eqnarray}
where $f$ is a random mean function, depending on the dose level $d$
and nanoparticle characteristics $\mathbf{x}_i$, and $\varepsilon_{ik}
\sim N(0, \sigma^2_d)$.
More precisely, $f$ is defined by a regression tree $\cT$ on the
predictor space and a functional model for dose-response curves in the
terminal nodes of $\cT$. Full details about the proposed mean structure
are described in the following section.

Given $f$, we assume that outcomes are independent across nanoparticles and,
for any nanoparticle $i$, 
$ \operatorname{Cov} (\varepsilon_{ik}(d), \varepsilon_{ik'}(d') ) =
\sigma^2 \varphi_D^{|d-d'|}$,
with $\varphi_D \in[0,1]$. In this setting, two outcomes associated
with the same nanoparticle at similar doses are assumed to be more
correlated than measurements taken at distant doses, for any replicate.
The major advantage of this assumption is related to a reduced
representation of a high-dimensional covariance matrix, which is now
fully characterized in terms of a $1$-dimensional variance parameter
$\sigma^2$ and a $1$-dimensional correlation~$\varphi_D$.

\subsection{Mean structure}\label{sectree}
The binary tree $\cT$ recursively splits the predictor space into two
subspaces, according to criteria of the form $x_{\cdot j} \leq a$ vs
$x_{\cdot j} >a$, for $a \in\mathbb{R}$ and $j \in\{ 1,\ldots,p\}$.
Each split defines two new nodes of the tree, corresponding to two
newly created subspaces of predictors. Let $n$ be the set of terminal
nodes of tree $\cT$.

We model the dose-response curves in each terminal node as a linear
combination of spline basis functions. Unlike parametric models such as
log-logistic, spline functions do not assume a particular shape for the
curve. This makes our model fully applicable to sub-lethal biological
assays, which are not expected to follow a sigmoidal dose-response dynamic.
However, if needed, the spline model can easily allow for possible
shape constraints, such as monotonicity, by using a modified basis
[\citet{RamseyMono}].
This flexibility makes the use of spline basis representations
potentially preferable to Gaussian process priors or similar smoothers.
A formal comparison is, however, outside the scope of this manuscript.
Our chosen functional representation is easily extended to
two-dimensional response surfaces (Section~\ref{sectiongeneralcase}).
Let $\mathcal{B}_1(\cdot),\ldots,\mathcal{B}_{m_D+\delta}(\cdot)$ denote
$m_D+\delta$ uniform B-spline basis functions of order $\delta$ on
$[0,D]$, with $m_D$ fixed knots. Following \citet{eilers}, we
avoid choosing the location of spline interior knots by deliberately
overfitting curves with a number of knots coinciding with the
dose-escalation grid. Adaptive smoothness is determined by using a
penalty on adjacent coefficients, via a smoothing prior that will be
presented in Section~\ref{sectionprior}.

If $\mathbf{x}_i$ is in the subset corresponding to the $r$th
terminal node of $\cT$,
$f(\mathbf{x}_i,d) = \sum_{\ell=1}^{m_D+\delta} \beta_{r \ell} \mathcal
{B}_\ell(d)$.
We will denote with $\bolds{\beta}_{r} = ( \beta_{r1}, \ldots, \beta_{r
m_D+\delta} )'$ the vector of splines coefficients defining the
expected dose-response trajectory in the $r$th terminal node.
Furthermore, we let $\bolds{\beta}$ define the random set of spline
coefficients, including $\bolds{\beta}_r$ from all terminal nodes $(r =
1,\ldots,n)$.
The Bayesian model is completed by prior distributions on $\cT$, $\bolds
{\beta}$, $\sigma^2$ and $\varphi_D$.

\section{Prior model}\label{sectionprior}
We first introduce the general dependence structure of the prior,
before describing each parameter's prior distribution. We follow
\citet{chipman97}, and assume that the tree is independent of
variance components $\sigma^2$ and $\varphi_D$:
%
\begin{equation}
\qquad p \bigl(\cT, \bolds{\beta}, \sigma^2, \varphi_D
\bigr) = p(\cT,\bolds{\beta}) p \bigl(\sigma^2 \bigr) p(
\varphi_D)
= p(\bolds{\beta}\mid\cT)p(\cT) p \bigl(
\sigma^2 \bigr) p( \varphi_D). \label{eqprior}
\end{equation}
Moreover, conditionally on $\cT$, terminal node parameters are assumed
independent: $p( \bolds{\beta} \mid T) = \prod_{r=1}^{n} p(\bolds{\beta
}_{r} \mid\cT)$.
Therefore, the prior is fully determined by a tree prior $p(\cT)$,
terminal node parameters priors $p(\bolds{\beta}_{r}\mid\cT)$, and
variance parameters priors $p(\sigma^2)$ and $p(\varphi_D)$.

\subsection{Tree prior}
The tree prior $p(\cT)$ is implicitly described by the stochastic
tree-generating process of \citet{chipman97}, where each new tree
is generated according to the following: (i) the probability for
a node at depth $q$ to be nonterminal, given by $\alpha(1+q)^{-\nu}$,
$(q=1,2,\ldots)$, (ii) the probability for a node to split at a
predictor $x_{\cdot j}$, $(j=1,\ldots, p)$, given by the discrete
uniform distribution on the set of available predictors, and
(iii) given the predictor $x_{\cdot j}$, the probability for a node
to split at a value $a$, given by the discrete uniform distribution on
the set of available splitting values.
Probability (i)
is a decreasing function of $q$, making deeper nodes less likely to
split and favoring ``bushy'' trees. \citet{chipman97} give
guidelines to choose parameters $\alpha$ and $\nu$ by plotting the
marginal prior distribution of the number of terminal nodes. In (ii)
and (iii),
predictors and splits are available if they lead to nonempty child nodes.

\subsection{Terminal node splines coefficients prior}
We follow \citet{lang} and consider a conditionally conjugate
P-spline prior:
$\bolds{\beta}_{r} \mid\cT, \tau^2 \propto \exp(- \frac{1}{2 \tau^2}
\bolds{\beta}_{r}' K_\beta\bolds{\beta}_{r} )$,
where $\tau^2$ is an additional smoothing variance parameter and
%
\begin{eqnarray}
K_\beta& = & \pmatrix{ 1 & -1 & & &
\cr
-1 & 2 & -1 & &
\cr
& \ddots&
\ddots& \ddots&
\cr
& & -1 & 2 & -1
\cr
& & & -1 & 1}\label{penmatrix}
\end{eqnarray}
is a penalty matrix of size $(m_D+\delta) \times(m_D+\delta)$,
corresponding to a first order random walk. Note that this prior is
improper, as the matrix $K_\beta$ is not of full rank. In order to work
with a proper prior in a model comparison setting, we replace the first
and last element of the diagonal with $1 + \eta$, where $\eta$ is a
small constant. The model is completed by assigning a conjugate
Inverse--Gamma hyperprior to the smoothing parameter $\tau^2 \mid\cT
\sim \mathrm{IG}(a_{\tau},b_{\tau})$.

\subsection{Variance components priors}\label{sec3.3}
We assume $\sigma^2 \sim \mathrm{IG}(a_\sigma, b_\sigma)$. For $\varphi_D$, we
choose the conjugate prior described by \citet{rowe} for
autoregressive covariance matrices, with truncated support on $[0,1]$.
Let $0=d_1<\cdots<d_{n_D}=D$ be the dose-escalation design sequence:
%
\begin{eqnarray}
\qquad p(\varphi_D) & \propto& \bigl( 1 - \varphi_D^2
\bigr)^{-(n_D -1)/2} \exp\biggl( - \frac{\lambda_{01} - \varphi
_D \lambda_{02} + \varphi_D^2 \lambda_{03}}{2 ( 1 - \varphi_D^2 )}
\biggr)
\mathbb{I}_{\varphi_D \in[0,1]}, \label{distphi}
\end{eqnarray}
where $\mathbb{I}$ is the indicator function, $\Lambda= (\Lambda
_{vv'})_{1 \leq v,v' \leq n_D}$ is a hyperparameter matrix, and
$(\lambda_{01}, \lambda_{02}, \lambda_{03})$ are defined through its
diagonal, subdiagonal, and superdiagonal elements as
follows: $ \lambda_{01} = \sum_{v=1}^{n_D} \Lambda_{vv}$,
$\lambda_{02} = \sum_{v=1}^{n_D-1} ( \Lambda_{vv+1}+ \Lambda_{v+1v} )$,
$\lambda_{03} = \sum_{d=2}^{n_D -1} \Lambda_{vv}$.
In practice, we choose $\Lambda= Id_{n_D}$, the identity matrix of
size $n_D \times n_D$, to put more weight on low values of $\varphi_D$
and assume weak prior correlations between responses at different
doses. This last distribution completes the prior model. We now turn to
posterior inference on parameters, given the observations.

\section{Posterior inference through MCMC simulation}\label{sectionalgo}

We are interested in the posterior distribution
$ p ( \cT,\bolds{\beta}, \sigma^2, \varphi_D, \tau^2 \mid\mathbf{y} )$.
The rest of this section describes a Markov chain Monte Carlo algorithm
for sampling from this distribution, as the number of potential trees
prevents direct calculations. Our Gibbs sampler is adapted from the
algorithms of \citet{chipman97} and \citet{TGP}, with changes
due to the specific structure of our model.

At each iteration, the algorithm performs a joint update of $(\cT,
\bolds{\beta})$, conditionally on the rest of the parameters, followed
by standard Gibbs component-wise updates of each variance parameter.
The joint tree and terminal nodes' spline coefficients update is
decomposed into
%
\begin{eqnarray}
&&\cT \mid \mathbf{y}, \sigma^2, \varphi_D,
\tau^2;\qquad\mbox{followed by } \label{eqtr}
\\
&&\bolds{\beta}_r  \mid \cT, \mathbf{y}, \sigma^2,
\varphi_D, \tau^2;\qquad\mbox{for } r \in\{1,\ldots,n\}.
\label{eqbeta}
\end{eqnarray}

The draw of $\cT$ in (\ref{eqtr}) is performed by the
Metropolis--Hastings algorithm of \citet{chipman97}, which
simulates a Markov chain of trees that converges to the posterior
distribution $p(\cT\mid\mathbf{y}, \sigma^2, \varphi_D, \tau^2)$. The
proposal density suggests a new tree based on four moves: grow a
terminal node, prune a pair of terminal nodes, change the split rule of
an internal node, and swap the splits of an internal node and one of
its children's.

The target distribution can be decomposed as follows:
%
\begin{eqnarray}
& & p \bigl(\cT\mid\mathbf{y}, \sigma^2, \varphi_D,
\tau^2 \bigr)
\nonumber\\[-8pt]\\[-8pt]\nonumber
&&\qquad \propto p(\cT) \int p \bigl(\mathbf{y} \mid\bolds{\beta
}, \cT, \sigma^2, \varphi_D, \tau^2 \bigr) p
\bigl( \bolds{\beta} \mid\cT, \sigma^2, \varphi_D,
\tau^2 \bigr) \,d \bolds{\beta}.
\end{eqnarray}
The expression for the integral above is given in \citet{supp}, in
a closed form by conjugacy of the prior on $\bolds{\beta} = \{\bolds
{\beta}_1,\ldots,\bolds{\beta}_n\}$. Therefore, the draw of $\cT$ in~(\ref{eqtr}) does not require a reversible-jump procedure for spaces of
varying dimensions, even if nodes are added or deleted.
The proposal density of the Metropolis--Hastings algorithm can be
conveniently coupled with $p(\cT)$ to simplify calculations [\citet
{chipman03}]. Full conditional distributions for $\bolds{\beta}_1,\ldots
,\bolds{\beta}_n$ in (\ref{eqbeta}) and variance parameters $\sigma^2$,
$\varphi_D$ and $\tau^2$ are available in \citet{supp}.

Given posterior samples, predictive statistics are easily obtained via
Monte Carlo simulation of $p(\mathbf{y}_i^*\mid\mathbf{y})$, for
$i=1,\ldots,I$. More precisely, let $\mathbf{x}_i^* = \mathbf{x}_i$.
At each iteration $\ell=1,\ldots,N$, the MCMC\vspace*{1pt} algorithm performs a draw
from $ p ( \cT,\bolds{\beta}, \sigma^2, \varphi_D, \tau^2 \mid\mathbf
{y} )$, followed by a draw of $\mathbf{y}_i^{(\ell)*}$ from the
multivariate normal distribution $p(\mathbf{y}_i^{(\ell)*}|\cT,\bolds
{\beta}, \sigma^2, \varphi_D, \tau^2)$. In our case studies
(Section~\ref{sectionappli}), for example, we compare posterior
summaries from the predictive distribution $p(y_{ik}^*(d) \mid\mathbf
{y})$ to observed dose-response data $y_{ik}(d)$.
We perform two series of posterior predictive checks: in the first one,
the generated predictive samples are conditioned on the full set of
dose-response curves, via the tree. The objective is to assess model
adequacy and calibration. The second series studies model prediction
accuracy using a leave-a-curve-out validation scheme, where each data
curve is compared to the corresponding predictive sample obtained by
fitting the tree on the remaining curves.

Posterior inference based on Monte Carlo samples is also used to derive
inferential summaries
about nontrivial functionals of the parameter/model space.
The marginal effect of a physico-chemical property $x_{.j}$ on the
response can be represented by the partial dependence function of
\citet{friedman}: let $x_{1j},\ldots,x_{Sj}$ be a grid of new
values for $x_{.j}$. Then the partial dependence function is
$ f(x_{sj},d,t) = ( \sum_{i=1}^{I} f ( (x_{i1},\ldots,x_{ij-1},x_{sj},
x_{ij+1},\ldots,x_{ip} ), d,t ) ) /I $,
where $x_{ij'}$ is the $i$th observation of $x_{.j'}$ in the data. For
all doses, plotting the average of this function over Monte Carlo draws
provides a visualization of the marginal effect of $x_{.j}$. This
partial dependence function can also be extended to account for the
joint marginal effect of two variables.

Similarly, posterior realizations $y|x$ can be used to report
importance scores for each variable. For all\vspace*{1pt} $j \in\{1,\ldots,P\}$,
$S_j = \frac{\var\{\mathbb{E}\{y|x_{.j}\}\}}{\var\{y\}}$ and $T_j =
\frac{\mathbb{E}\{ \var\{y|x_{.-j}\}\}}{ \var\{y\}}$
are the \emph{first-order} and \emph{total} sensitivity indices for
variable $x_{.j}$, and represent the main and total influence,
respectively, of this variable on the response [\citet{gramacy13}].
Unlike other metrics such as the variance reduction attributed to
splits on the variable, sensitivity indices are robust to leaf model
specifications and are therefore adapted for a dose-response leaf model.
Both indices are defined given an uncertainty distribution on the
inputs, usually the uniform distribution on the covariates space.
We follow \citet{gramacy13} and use a Monte Carlo scheme to
approximate $S_j$ and $T_j$, that accounts for unknown responses by
using predicted values for a Latin hypercube sampling design.

\section{Extending the model to two-dimensional toxicity profiles}\label{sectiongeneralcase}

More general exposure escalation protocols involve the observation of a
biological outcome $y$
in association with a prescription of dose escalation $d\in[0, D]$,
observed for a series of
exposure times $t\in[t_0, T]$. Letting $k$, $(k=1,\ldots,K)$ be a
replication index, we define $y_{ik}(d,t)$
as the outcome of interest, evaluated at dose $d$, time $t$ and extend
the model in (\ref{eqmodel1}):
$y_{ik}(d,t) = f( \mathbf{x}_i,d,t) + \varepsilon_{ik}(d,t)$,
where $f$ is a random mean response surface and $\varepsilon
_{ik}(d,t)\sim N(0, \sigma^2_{dt})$. To account for dependence between
doses and durations of exposure, for each nanoparticle $i$, we assume
$\operatorname{Cov} ( \varepsilon_{ik}(d, t), \varepsilon_{ik'}(d', t')
) = \sigma^2 \varphi_D^{|d-d'|} \varphi_T^{|t-t'|}$,
where $\varphi_D\in[0,1]$ and $\varphi_T \in[0,1]$ are autocorrelation
parameters.

The response surface $f$ in the terminal nodes of $\cT$ is modeled by a
tensor product of two one-dimensional P-splines [\citet{lang}].
Let $\mathcal{B}_1(\cdot),\ldots,\mathcal{B}_{m_D+\delta}(\cdot)$ defined as in
Section~2.2 and $\mathcal{B}_1(\cdot),\ldots,\mathcal{B}_{m_T+\zeta}(\cdot)$
denote $m_T+\zeta$ B-spline basis functions of order $\zeta$ on
$[t_0,T]$, with $m_T$ fixed knots.
Then, if $\mathbf{x}_i$ is in the subset corresponding to the $r$th
terminal node of $T_j$,
$ f( \mathbf{x}_i,d,t) = \sum_{\ell=1}^{m_D+\delta} \sum
_{m=1}^{m_T+\zeta} \beta_{r \ell m} \mathcal{B}_\ell(d) \mathcal{B}_m(t)$,
where $\bolds{\beta}_{r} = ( \beta_{r11}, \ldots,\break  \beta_{r (m_D+\delta)
(m_T+\zeta)} )'$ is a vector of spline coefficients associated to the
$r$th terminal node.

The prior model has the same global dependence structure as in
Section~\ref{sectionprior}, but now includes an additional independent
term $\varphi_T$ for time-covariance. Let $t_0=t_1 <\cdots<t_{n_T}=T$
be the sequence of exposure times when toxicity was measured.
We adapt prior (\ref{distphi}) to preserve conjugacy and introduce a
similar distribution for $\varphi_T$:
%
\begin{eqnarray}
\qquad\quad p(\varphi_D) &\propto& \bigl( 1 - \varphi_D^2
\bigr)^{-(n_T(n_D-1))/2} \exp\biggl( - \frac{\lambda_{01} -
\varphi_D \lambda_{02} + \varphi_D^2 \lambda_{03}}{2 ( 1 - \varphi_D^2
)} \biggr)
\mathbb{I}_{\varphi_D \in[0,1]},
\\
p(\varphi_T) &\propto&\bigl( 1 -
\varphi_T^2 \bigr)^{-(n_D(n_T-1))/2} \exp\biggl( -
\frac{\gamma_{01} - \varphi_T \gamma_{02} + \varphi_T^2 \gamma_{03}}{2
( 1 - \varphi_T^2 )} \biggr) \mathbb{I}_{\varphi_T \in[0,1]},
\end{eqnarray}
%
where $\gamma_{01}$, $\gamma_{02}$ and $\gamma_{03}$ are obtained by
summing elements of the diagonal, subdiagonal, and superdiagonal of
matrix parameter prior $\Gamma$, constructed following the guidelines
introduced in Section~\ref{sec3.3}.
For the terminal nodes' spline coefficient priors, we use a spatial
extension of \citet{besag}, a first order random walk prior based
on the four nearest neighbours of splines coefficients, with
appropriate changes for corners\vspace*{1pt} and edges:
$\bolds{\beta}_{r} | \cT, \tau^2 \propto\exp(- \frac{1}{2 \tau^2}
\bolds{\beta}_{r}' K_{\beta} \bolds{\beta}_{r} )$,
where $K_\beta$ is a penalty band matrix of size $(m_D+\delta)(m_T+\zeta
) \times(m_D+\delta)(m_T+\zeta)$, which extends matrix (\ref
{penmatrix}) to the two-dimensional case. For posterior inference, we
add a step to generate $\varphi_T$ in the Gibbs sampler of Section~\ref
{sectionalgo}.

\section{Applications}
\label{sectionappli}
A simulation study to assess model performance
is described in \citet{supp}. In the rest of this section we
illustrate our approach with experimental results from a case study
reported by \citet{zhang}, measuring toxicity of 24 metal oxides
on human bronchial epithelial (BEAS-2B) cells.

\subsection{Case studies background}
\begin{figure}

\includegraphics{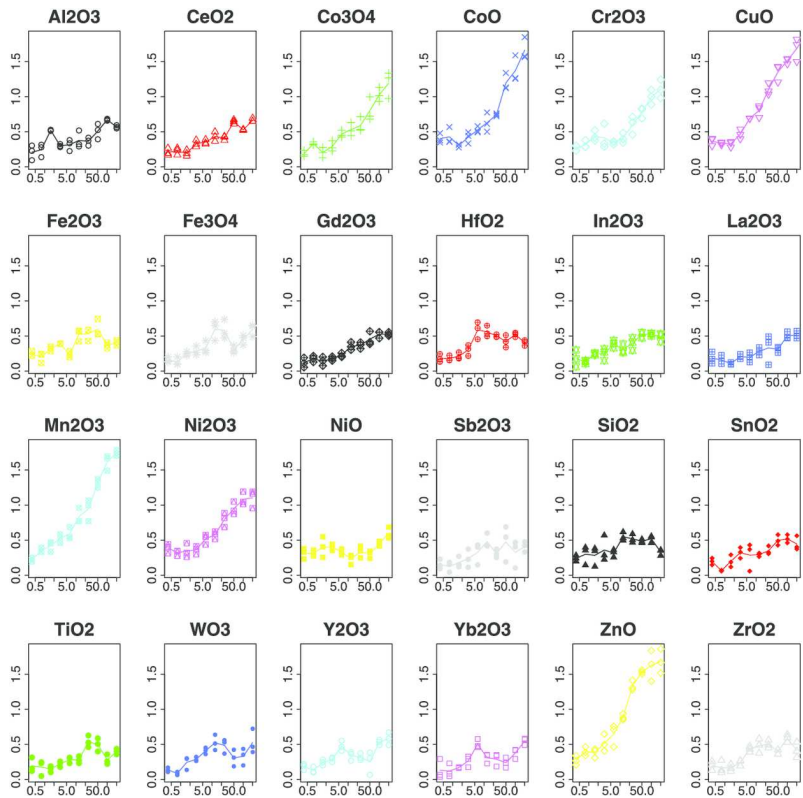}

\caption{Dose-response curves for LDH assay.}
\label{fig1}
\end{figure}

\begin{figure}

\includegraphics{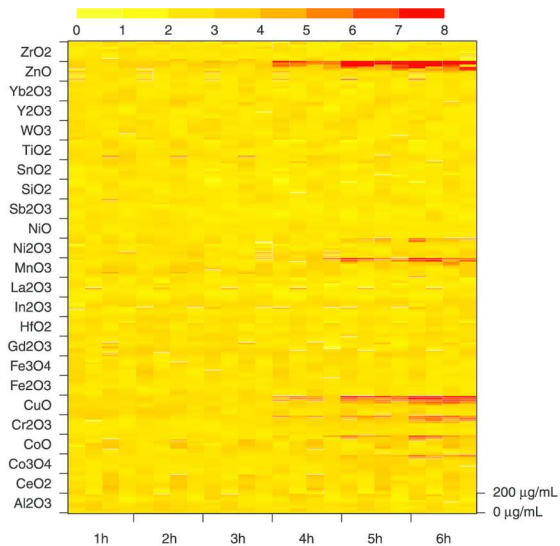}

\caption{Heatmap for PI fluorescence assay, color-coded from light
(low) to dark (high). Each row corresponds to a nanoparticle at one
dose across 6 times (1 to 6~h) and 4 replicates. For each nanoparticle,
there are 11 rows, one for each dose (0 to 200~\textmu g${}/{}$ml), arranged from
bottom to top.}\label{fig2}
\end{figure}

After 24~h, Lactate Dehydrogenase (LDH) release was used to measure the
death rate of cells exposed to eleven doses of metal oxides (from 0 to
200~\textmu g${}/{}$ml), evenly spaced on the logarithmic scale. Cell death is
commonly used to screen for ENM cytotoxicity without reference to a
specific mechanism.
Figure~\ref{fig1} shows the LDH dose-responses curves for the 24 metal
oxide nanoparticles.
In a second assay, Propidium Iodide (PI) fluorescence was used to
indicate the percentage of cells experiencing oxidative stress through
cellular surface membrane permeability,
across the same ten doses and after six times of exposure (from 1 to
6~h, at every hour).
Figure~\ref{fig2} shows a heatmap representation for the PI assay, for
all metal oxides, doses, times, and replicates, where responses are
color-coded from light (low) to dark (high).
In both assays, seven metal oxides (Co$_3$O$_4$, CoO, Cr$_2$O$_3$, CuO,
Mn$_2$O$_3$, Ni$_2$O$_3$ and ZnO) display a notable rise for the
higher doses, suggesting toxicity.

%

%

All metal oxides are characterized by six physico-chemical properties
of potential interest to explain toxicity profiles: nanoparticle size
in media, a measure of the crystalline structure (b(\r{A})), lattice
energy ($\Delta H_{\mathrm{lattice}}$), which measures the strength of the bonds
in the nanoparticles, the enthalpy of formation ($\Delta H_{Me^{n+}}$),
which is a combined measure of the energy required to convert a solid
to a gas and the energy required to remove $n$ electrons from that gas,
metal dissolution rate, and conduction band energy (the energy to free
electrons from binding with atoms).

In our analysis,
we use cubic splines, that is, $\delta= \zeta=4$, and place interior
knots at each intermediate dose from 0.39 to 100~\textmu g${}/{}$ml. Therefore,
$n_D = m_D+2$ and $n_T = m_T+2$.
For the tree prior, we adopt the default choice of \citet
{chipman97}, $(\alpha,\nu) =(0.95,2)$, which puts more weight on trees
of size 2 or 3. We place\vspace*{1pt} relatively diffuse priors $\operatorname{Gamma}(1,1)$ on
precision parameters $1/\tau^2$ and $1/\sigma^2$. We choose $\Lambda=
Id_{n_D}$ and $\Gamma= Id_{n_T}$,
assuming no prior correlations between measurements at different doses
and times. Finally, moves ``Grow,'' ``Prune,'' ``Change'' and ``Swap''
of the Metropolis--Hastings tree-generating algorithm have probabilities
$0.1$, $0.1$, $0.6$ and $0.2$, respectively.
We used a total of 160,000 iterations. After discarding 80,000 iterations
for burn-in, the remaining samples for estimation were thinned to save
computer storage. The rest of this section shows the results obtained
on LDH and PI assays.


\subsection{LDH dose-escalation assay}
\begin{figure}

\includegraphics{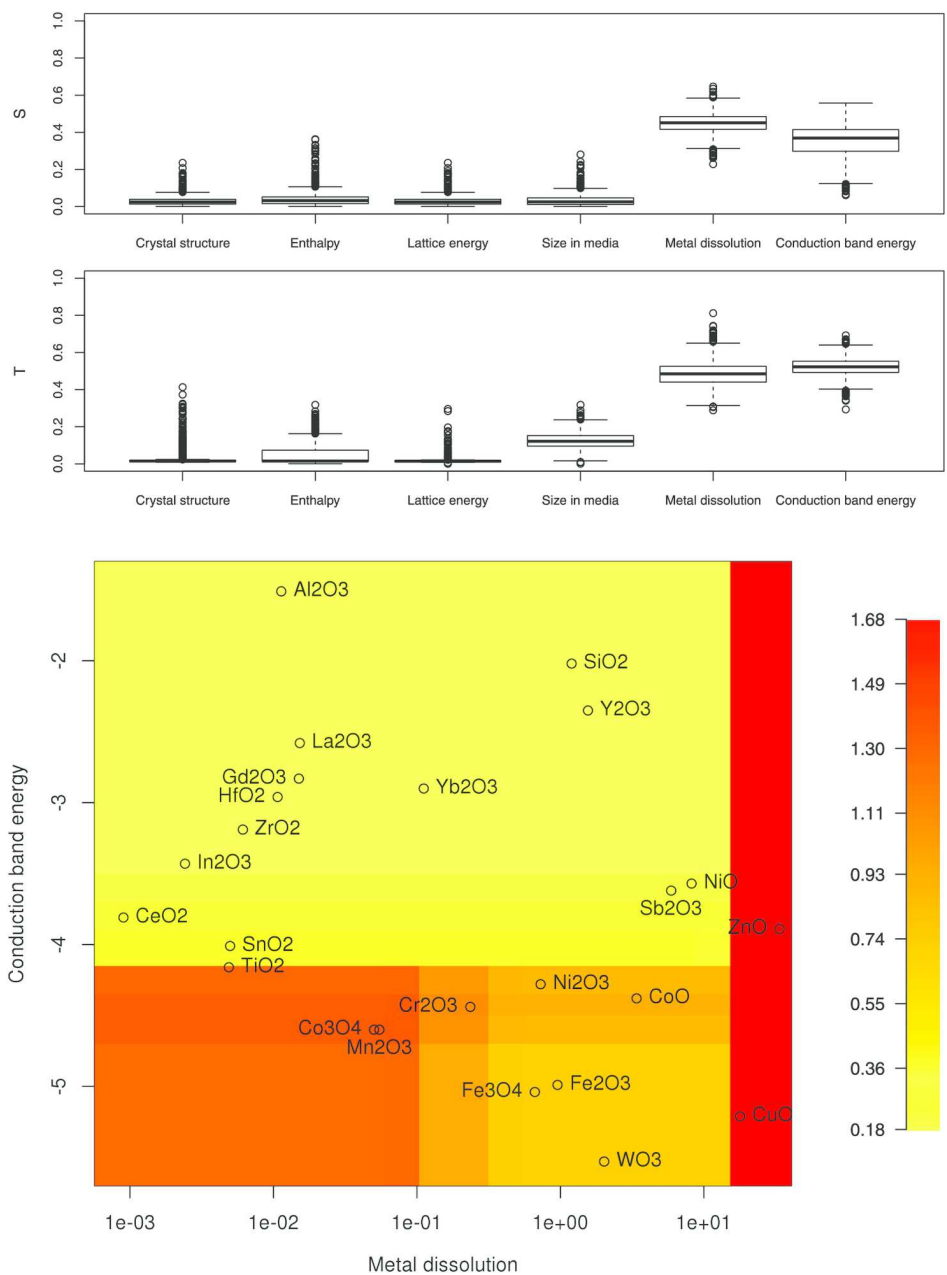}

\caption{\emph{LDH assay.} {(Top)} First order (S) and total (T)
sensitivity indices for the six physio-chemical properties in the LDH
assay. {(Bottom)} 2-dimensional partial dependence function for
marginal effect of metal dissolution (log scale) and conduction band
energy in the LDH assay at 200~\textmu g${}/{}$ml. The toxicity response is
color-coded from light (low) to dark (high). The figure also shows the
projections of the 24 metaloxides in this subspace.}\label{fig4}
\end{figure}

Figure~\ref{fig4} {(top)} shows both sensitivity indices described
in Section~\ref{sectionalgo} for the six physico-chemical properties.
Figure~\ref{fig4} {(bottom)} shows the combined marginal effect of
conduction band energy and dissolution on LDH, obtained with the
partial dependence function of \citet{friedman}, and color-coded
from light (low) to dark (high), for dose 200~\textmu g${}/{}$ml. The tree
isolates a first region of high toxicity, corresponding to ENM with
high dissolution rates (ZnO and CuO). This region corresponds to the
first mechanism of toxicity identified by \citet{zhang}: highly
soluble metal oxides, such as ZnO and CuO, are more likely to release
metal ions and disturb the cellular state.
A second region of toxicity on Figure~\ref{fig4} {(left)} includes
metal oxides Co$_3$O$_4$, CoO, Cr$_2$O$_3$, Mn$_2$O$_3$ and
Ni$_2$O$_3$, with Ec values ranging from $-$4.33 eV for Mn$_2$O$_3$ to
$-$4.59 eV for Ni$_2$O$_3$. This region matches the second mechanism for
toxicity described by \citet{zhang}: the overlap of the conduction
band energy of the metal oxides with the biological redox potential of
cells, ranging from $-$4.12 to $-$4.84~eV. When these two energy levels are
alike, transfer of electrons from metal oxides to cells is facilitated,
disturbing the intracellular state. Note that Figure~\ref{fig4}
{(bottom)} also shows an additional split that isolates Mn$_2$O$_3$,
whose toxicity for the LDH assay is more comparable to ZnO and CuO (see
Figure~\ref{fig1}).
Similar figures for other doses are included in \citet{supp}.
The LDH assay illustrates how threshold effects and interactions of
physico-chemical properties are accurately captured by a tree structure.
%

\begin{figure}

\includegraphics{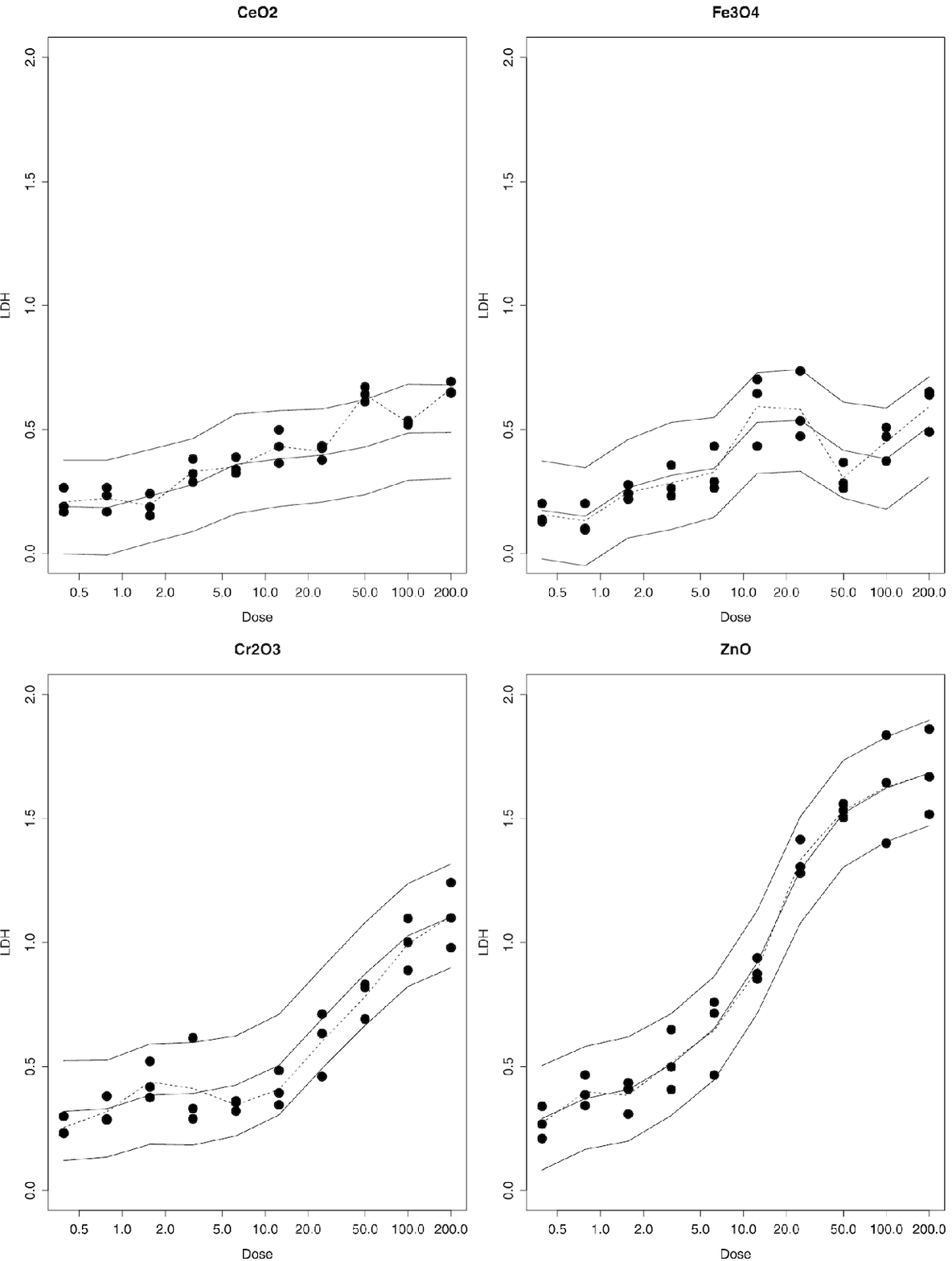}

\caption{Posterior predictive curves for CeO$_2$, Fe$_3$O$_4$,
Cr$_2$O$_3$ and ZnO. The points are the observed replicates and the
dashed line is the average observed response. The expected posterior
predictive curve and $90\%$ interval are in solid lines.}\label{fig3}
\end{figure}

We perform posterior predictive checks for model fitting. Figure~\ref
{fig3} shows the expected posterior predictive dose-response curves for
two nontoxic metal oxides (CeO$_2$ and Fe$_3$O$_4$) and two toxic ones
(Cr$_2$O$_3$ and ZnO), with the associated $90\%$ intervals.
All four intervals provide good coverage for the original data. The
other 20 curves exhibit similar behavior and can be found in \citet{supp}.
We also study the prediction accuracy of the model using a
leave-a-curve-out validation framework. Results for CeO$_2$,
Fe$_3$O$_4$, Cr$_2$O$_3$ and ZnO are presented
in \citet{supp}.
While leave-one-out predictions recover general trends, in some cases
we observe suboptimal coverage, especially in sparse areas of the
physico-chemical spectrum. For example, nanoparticles ZnO and CuO alone
determine tree splits on the metal dissolution parameter and, once
removed, cannot be accurately predicted by the model.

%

Finally, the proposed methodology is compared for validation to the\break 
Bayesian Additive Regression Trees (BART) method of \citet{bart},
a sum-of-tree extension of \citet{chipman97}, with the R package
``BayesTree'' [\citet{BayesTree}]. As BART model one-dimensional
responses, we use the area under the LDH curves (AUC) as the dependent
variable. In \citet{bart}, the proportion of all splitting rules
attributed to a variable at each draw on all trees, averaged over all
iterations, is proposed as a measure of variable importance, when the
number of trees is small.
Results are presented in \citet{supp}. Variable importance scores
and marginal effects from BART are similar to those obtained with our
method and confirm that the AUC is an accurate summary for toxicity for
the LDH assay.
The first advantage of using a dose-response leaf model instead of the
AUC is that we avoid preliminary assessment of the data for choosing a
summary over another: toxicologists usually report several toxicity
parameters (EC50, slope), as they may convey different information. The
second advantage is better understood from a predictive perspective, as
our model allows for full dose-response dynamics instead of the AUC.
A comparison with the Treed Gaussian Process, using the \verb'R'
package \verb'tgp' [\citet{gramacy}], is also included in
\citet{supp}. After tuning \verb'tgp' to
forbid splitting on dose (\verb'basemax', \verb'splitmin'),
we can indeed reproduce the essential structure of our model using this
well-tested \verb'R' library. Our findings proved to be robust to
differing details in the prior specification, as the model fit with
\verb'tgp' also captures the marginal effects of the predictors metal
dissolution and conduction band energy on toxicity.

\subsection{PI general exposure assay}

%
\begin{figure}

\includegraphics{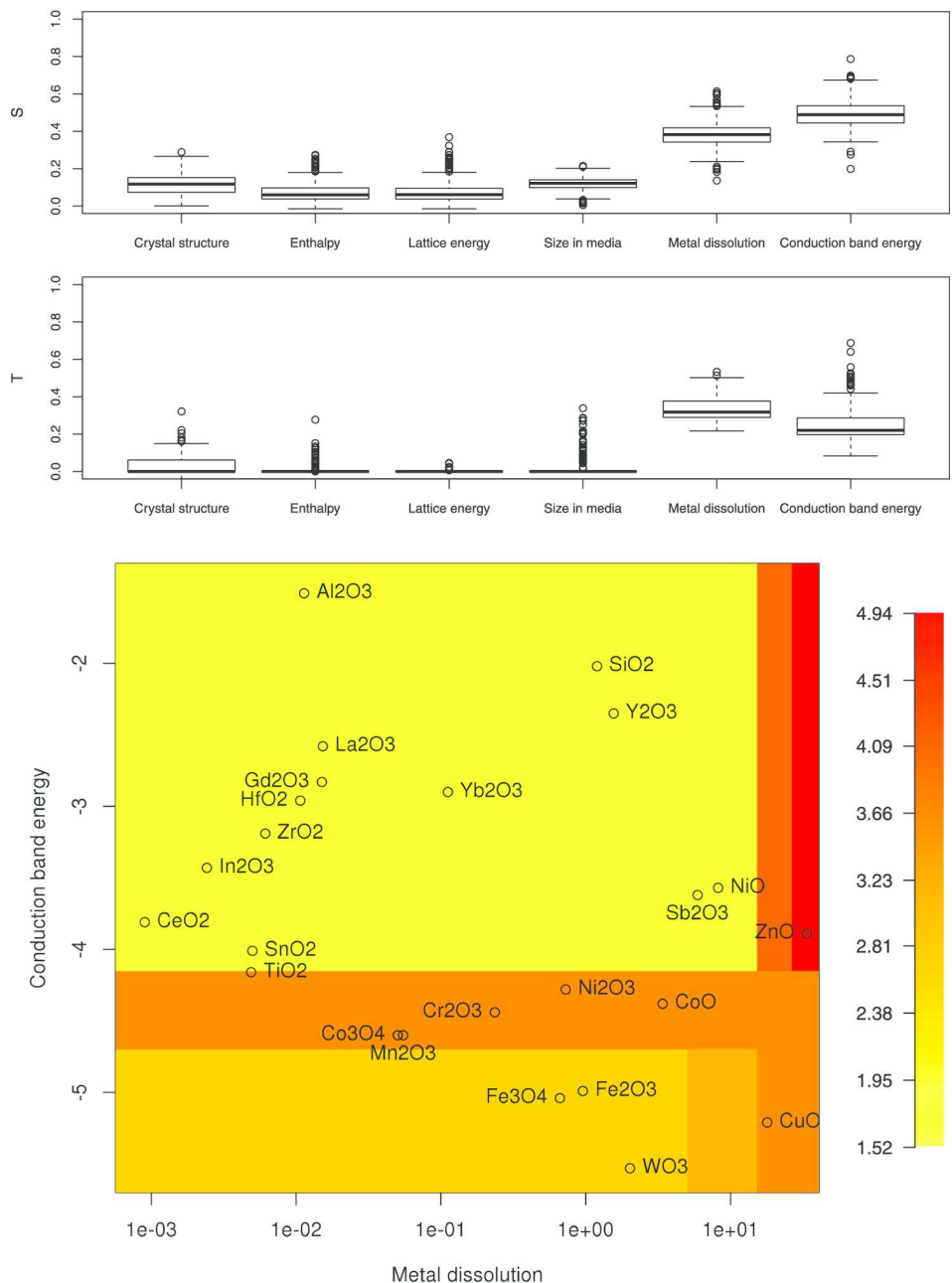}

\caption{\emph{PI uptake.} {(Top)} First order (S) and total
(T) sensitivity indices for the six physio-chemical properties in the
PI assay. {(Bottom)} 2-dimensional partial dependence function for
marginal effect of metal dissolution (log scale) and conduction band
energy in the PI assay at 200~\textmu g${}/{}$ml and 6~h. The toxicity response
is color-coded from light (low) to dark (high). The figure also shows
the projections of the 24 metaloxides in this subspace.}
\label{fig9}
\end{figure}

%
\begin{figure}

\includegraphics{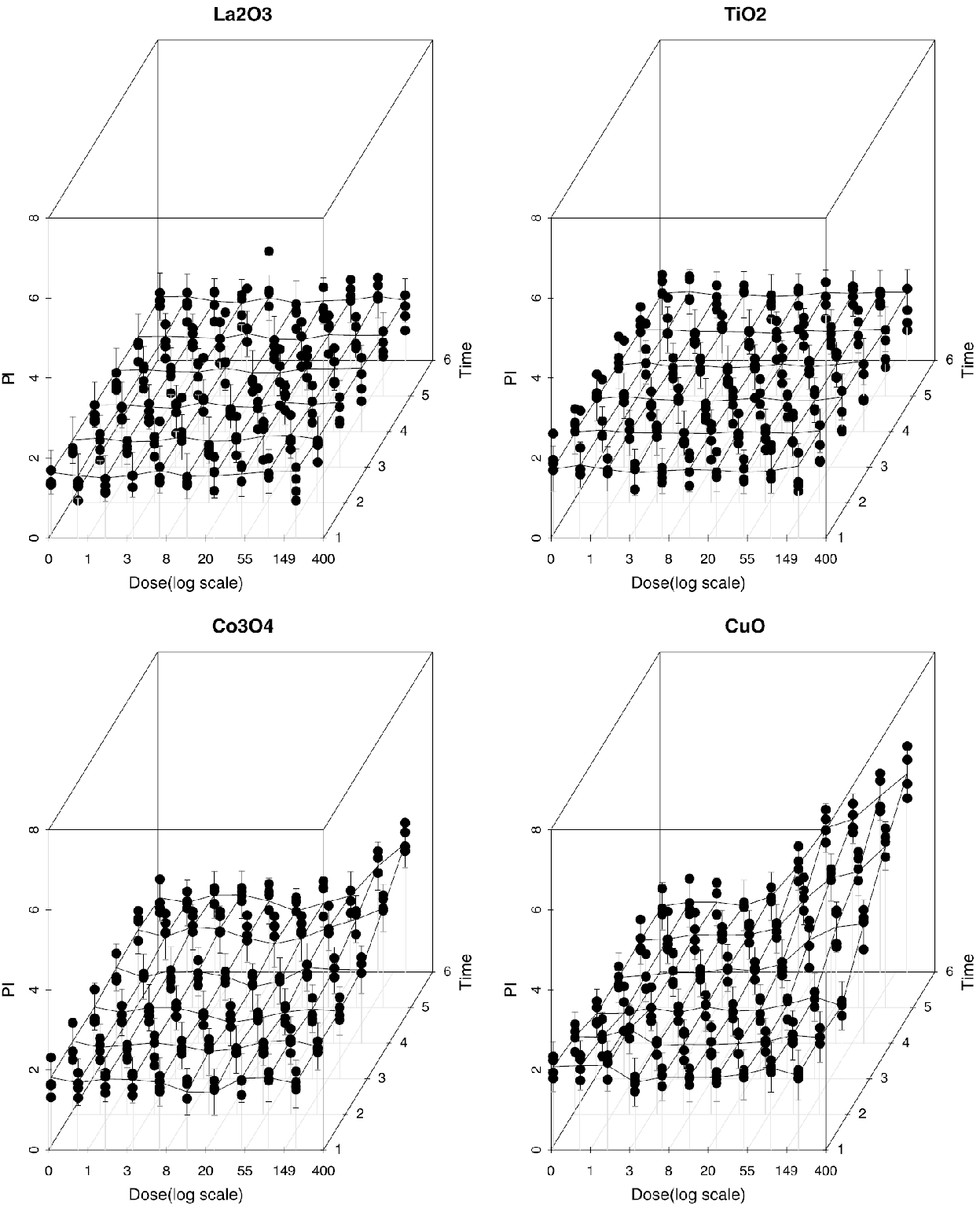}

\caption{Posterior predictive surfaces for La$_2$O$_3$, TiO$_2$,
Co$_3$O$_4$ and CuO. The solid line is the expected posterior
predictive surface with the associated $90\%$ interval. The points are
the observed data replicates.}
\label{fig8}
\end{figure}

Figure~\ref{fig9} {(top)} shows the variable sensitivity indices of
the six physico-chemical properties.
Figure~\ref{fig9} {(bottom)} illustrates the marginal effect of
both conduction band energy and dissolution on membrane damage,
calculated with the partial dependence function, and color-coded from
light to dark, for dose 200~\textmu g${}/{}$ml and time 6~h. The tree model
for PI assay also identifies the two areas of toxicity indicated in
\citet{zhang}, corresponding to highly soluble metal oxides and
nanoparticles whose conduction band energy overlaps with cellular redox
potential range.
Additional figures for marginal effect of conduction band energy and
metal dissolution, for all doses and times, are included in \citet{supp}.
The similarity of variable importance scores and marginal effect of
conduction band energy and dissolution obtained for LDH and PI assays
indicates a strong correlation between these assays for nanoparticle
toxicity assessment, as noted by \citet{zhang}.
Figure~\ref{fig8} illustrates the posterior predictive $90\%$ surface
intervals for two nontoxic metal oxides (La$_2$O$_3$ and TiO$_2$) and
two toxic ones (Co$_3$O$_4$ and CuO), showing good posterior coverage
over all doses and times of exposure.
Similar surfaces for the other 20 metal oxides are plotted in
\citet{supp}.
Leave-a-surface-out predictions for La$_2$O$_3$, TiO$_2$, Co$_3$O$_4$,
and CuO are presented
in the appendix and show the limitations of the model for prediction
when extrapolating to sparse areas of the covariate space, similar to
what we observed in the LDH assay.

\section{Discussion}
\label{sectiondiscu}
We propose a Bayesian regression tree model to define relationships
between physico-chemical properties of engineered nanomaterials and
their functional toxicity profiles in dose-escalation assays.
As demonstrated by the case studies, the tree structure is adapted to
account for flexible models of structure-activity relationships, such
as threshold effects and interactions.
The proposed model integrates information across all doses and
replicates, and therefore is adapted to small sample sizes usually
found in nanotoxicology data sets.
Monte Carlo integration over the model space provides straightforward
inference on nontrivial functionals of parameters of interest and
prediction of full dose-response curves from nanoparticle characteristics.
The smoothing splines representation allows for easy extension of the
model to two-dimensional toxicity profiles of general exposure
escalation assays as well as for modeling sub-lethal outcomes.

The convergence of Bayesian tree models should be carefully assessed
for all applications of the proposed methodology. The four moves of the
Metropolis--Hastings algorithm of \citet{chipman97} work well in
our simulations and case studies, however, other applications might
require additional moves to move faster through the tree space and
improve convergence [see, e.g., \citet{TGP,wu}].
As illustrated in Section~\ref{sectionappli}, another potential pitfall
of the model is its predictive performance for sparsely explored
nanoparticle characteristics. This issue is not specific to our model and
possible improvements would be obtained by combining multiple studies
in a meta-analysis framework, with the appropriate adjustments for data
heterogeneity or formalizing explicit prior knowledge about hazardous
nanoparticle properties. 

As seen in the case study for cell death and cellular membrane
permeability, different toxicity mechanisms can be closely related.
Therefore, an important opportunity for model extensions would be to
combine different biological assays in a single analysis, the final
goal being that of understanding if nanoparticles physical and chemical
properties have
a differential effect on different cellular injury pathways. This would
require more sophisticated modeling strategies that will be more likely
to be useful if technological advances will allow for feasible
screening of much larger nanomaterial libraries.

\section*{Acknowledgments}
Any opinions, findings, conclusions or
recommendations
expressed herein are those of the author(s) and do not necessarily
reflect the views of the
National Science Foundation or the Environmental Protection Agency.
This work has not been subjected to an EPA peer and policy review.

\begin{supplement}[id=suppA]
\stitle{\hspace*{-2pt}Additional results for online publication}
\slink[doi]{10.1214/14-AOAS797SUPPA} 
\sdatatype{.pdf}
\sfilename{aoas797\_suppa.pdf}
\sdescription{This appendix provides full conditional distributions and additional experimental results.}
\end{supplement}

\begin{supplement}[id=suppB]
\stitle{\hspace*{-2pt}Code}
\slink[doi]{10.1214/14-AOAS797SUPPB} 
\sdatatype{.zip}
\sfilename{aoas797\_suppb.zip}
\sdescription{This folder contains a C++ implementation of the algorithm.}
\end{supplement}

%

\printaddresses
\end{document}